\documentclass[12pt]{iopart}

\usepackage{multicol}
\usepackage{graphicx}
\usepackage{booktabs}
\usepackage{amssymb,bm,mathrsfs,bbm,amscd}
\usepackage{lastpage}
\usepackage{CJK}

\def\bd{\begin{document}} \def\ed{\end{document}}
\def\bmp{\begin{minipage}} \def\emp{\end{minipage}}
\def\bcc{\begin{center}} \def\ecc{\end{center}}     \def\npg{\newpage}
\def\beq{\begin{equation}} \def\eeq{\end{equation}} \def\hph{\hphantom}
 \def\r#1{$^{[#1]}$}
\def\n{\noindent} \def\ni{\noindent} \def\pa{\parindent}
\def\hs{\hskip} \def\vs{\vskip} \def\hf{\hfill} \def\ej{\vfill\eject}
\def\cl{\centerline} \def\ob{\obeylines}  \def\ls{\leftskip}
\def\underbar#1{$\setbox0=\hbox{#1} \dp0=1.5pt \mathsurround=0pt
   \underline{\box0}$}   \def\ub{\underbar}    \def\ul{\underline}
\def\f{\left} \def\g{\right} \def\e{{\rm e}} \def\o{\over} \def\d{{\rm d}}
\def\vf{\varphi} \def\pl{\partial} \def\cov{{\rm cov}} \def\ch{{\rm ch}}
\def\la{\langle} \def\ra{\rangle} \def\EE{e$^+$e$^-$} \def\pt{p_{\rm T}}
\def\pti{p_{{\rm T},i}} \def\yti{y_{{\rm T},i}}
\def\ptj{p_{{\rm T},j}}\def\mt{m_{\rm T}} \def\yt{y_{\rm T}} \def\vt{v_{\rm T}}

\def\bitz{\begin{itemize}} \def\eitz{\end{itemize}}
\def\btbl{\begin{tabular}} \def\etbl{\end{tabular}}
\def\btbb{\begin{tabbing}} \def\etbb{\end{tabbing}}
\def\beqar{\begin{eqnarray}} \def\eeqar{\end{eqnarray}}
\def\\{\hfill\break} \def\dit{\item{-}} \def\i{\item}
\def\bbb{\begin{thebibliography}{9} \itemsep=-1mm}
\def\ebb{\end{thebibliography}} \def\bb{\bibitem}
\def\bpic{\begin{picture}(260,240)} \def\epic{\end{picture}}
\def\akgt{\cl{\bf ACKNOWLEDGMENTS}}
\def\fgn{\noindent{\bf\large\bf figure captions}}

\def\lan{\langle}
\def\ran{\rangle}
\def\p{\pi}
\def\ifmath#1{\relax\ifmmode #1\else $#1$\fi}%
\def\rc{\ifmath{{\mathrm{c}}}}
\def\cut{\ifmath{{\mathrm{cut}}}}
\def\rF{\ifmath{{\mathrm{F}}}}
\def\rK{\ifmath{{\mathrm{K}}}}
\def\rp{\ifmath{{\mathrm{p}}}}
\def\rt{\ifmath{{\mathrm{t}}}}
\def\LAB{\ifmath{{\mathrm{LAB}}}}
\def\cut{\ifmath{{\mathrm{cut}}}}

\newcommand{\gguide}{{\it Preparing graphics for IOP journals}}
\begin{document}


\title[]{Influence of statistics on the measured moments of conserved quantities in relativistic heavy ion collisions}

\author{Lizhu Chen$^{1,2}$,  Zhiming Li$^2$, Xia Zhong$^1$, Yuncun He$^3$ and Yuanfang Wu$^{2,*}$}
\address{$^1$ School of Physics and Optoelectronic Engineering, Nanjing University of Information Science and Technology, Nanjing 210044, China}
\address{$^2$ Key Laboratory of Quark and Lepton Physics (MOE) and Institute of Particle Physics, Central China Normal University, Wuhan 430079, China}
\address{$^3$ Faculty of Physics and Electronic Technology, Hubei University, Wuhan 430079, China}
\address{$*$ Email: wuyf@phy.ccnu.edu.cn}

\begin{abstract}
We study statistics dependence of the probability distributions and the means of measured moments of conserved quantities, respectively. The required statistics of all interested moments and their products are estimated based on a simple simulation. We also explain why the measured moments are underestimated when the statistics are  insufficient.
With the statistics at RHIC/BES, the second and third order moments can be reliably obtained based on the
method of Centrality bin width correction (CBWC), which can not be applied for the fourth order moments at low energy.
With planning statistics at RHIC/BES II, and improved CBWC method, $\kappa\sigma^2$ in a finer centrality bin scale should be measurable. This will help us to understand  the current observation of energy and centrality dependence of high-order moments.

\end{abstract}


\section{Introduction}
\vspace{1mm}

The main goal of the RHIC Beam Energy Scan (BES) program is to search for the
Quantum Chromodynamics (QCD) critical point and phase boundary~\cite{STAR-BESI}. The calculations from lattice QCD and QCD based models have shown that the high-order moments of conserved quantities, such as net-baryon, net-charge and net-strangeness, are sensitive to
the critical fluctuations~\cite{QCD-1, QCD-2, QCD-3, QCD-4, QCD-5, QCD-6}. So the
behavior of high-order moments of conserved quantities are greatly interested in heavy ion collisions~\cite{Karsch-frozen-out-1, Karsch-frozen-out-2, Nu-science}.

However, the moments of conserved quantities are also affected by many non-critical backgrounds~\cite{koch-cpod-2014}, such as global baryon/charge conservation.  The moments are substantially suppressed if global conservation is taken into consideration~\cite{baryon-conservation}. In addition, artificial cuts of phase space in experimental analysis are not trivial~\cite{luo-1302}, such as the acceptance in rapidity and transverse momentum space, the selection of centrality of the collisions, and so on.


To reduce those non-critical background contributions, many new techniques are applied in moments analysis at RHIC/STAR~\cite{STAR-proton, STAR-charge}. The Delta theorem method is used for estimation of statistical uncertainties~\cite{delta-error}. The track efficiency correction~\cite{acceptance-koch} is firstly considered in event-by-event fluctuation.
The initial size fluctuation is one of the important non-critical effects. Although it has not been taken into consideration in most of the theoretical calculations~\cite{initial-1, QCD-Gupta, HRG-Karsch}, it exists in all event variables in experiments. As we know, the initial size is usually quantified by the centrality of the collisions, which is determined by the multiplicity ($N_{ch}$) of the final state. A centrality bin corresponds to a range of multiplicity. For a given centrality bin, i.e., a range of multiplicity, the initial size still fluctuates from event to event~\cite{impact}. In order to reduce this initial size fluctuation, it is proposed to calculate the moment in each of $N_{ch}$. The measured moment is averaged over all  $N_{ch}$ in a given centrality, where the average is weighted by the number of events in each of $N_{ch}$. This is so called  the method of  the  Centrality Bin Width Correction (CBWC)~\cite{cbwc-star}.

The CBWC method can only be reliably used if the statistics in each of $N_{ch}$ is large enough~\cite{statistics-lizhu}. It is shown that measured moments increase with statistics and finally saturate when statistics is sufficient. If the statistics is insufficient, the measured moments will be underestimated. In fact, it is already found that the fourth order moments would be underestimated at RHIC low energies.
In order to increase the statistics, it is suggested to calculate the moment at $\delta 1\%$ centrality bin width, instead of each of $N_{ch}$, in the CBWC method, i.e., so called the improved CBWC method~\cite{statistics-lizhu}. It can reduce the initial size fluctuations as well.


The published STAR results show $\kappa\sigma^2$ of net-proton multiplicity distributions in central collisions (0-5\% and 5-10\% centrality) is systematically lower than that in the other centralities~\cite{STAR-proton}.
The behavior of $\kappa\sigma^2$ in a  finer centrality bin (such as $\delta1\%$ centrality) should be further investigated.
At planning RHIC/BES II, the statistics will be increased to a few hundreds millions, see Table~\ref{statistics-BESII}~\cite{STAR-BESII}. With these large statistics, and improved CBWC method, it is possible to study the centrality dependence of high-order moments in a finer centrality bin scale, i.e., in each $\delta 1\%$ centrality bin width for central collisions (0-10\% centrality). If this can be done,
 the  centrality dependence of high-order moments can be understood better.

\begin{table}[htmp]
\centering
\begin {tabular}{|c|c|c|c|c|c|c|} \hline
Energy (GeV)  & 7.7  &  9.1 &  11.5 & 13.0 & 14.6 & 19.6\\
\hline
No. Events (Million)  & 100   & 160 & 230 & 250 & 300 & 400\\
\hline
\end{tabular}
\caption{The proposed statistics at RHIC/BES II.\label{statistics-BESII}}
\end{table}

In this paper,  we firstly study the statistics dependence of probability distribution of measured moments in section II. It is shown that when the probability distribution of measured moments
can be described by a normal one with fit goodness $\chi^2/ndf < 10$, the corresponding statistics is enough for the moment analysis. The required statistics in each of $N_{ch}$ or $\delta1\%$ centrality bin width for all interested moments and their product is estimated.
The statistics dependence of the means of measured moments are studied in section III. The required statistics is obtained when the mean of the measured moment is saturated to its expectation. By this way, the required statistics is lower than those obtained from probability distribution of measured moments. It also explains why the measured mean of moment is systematically smaller than its expectation when the statistics is short.

According to the statistics dependence of measured moments given in section II and III, the reliability of measured moments in current RHIC/BES I, and planed RHIC/BES II are discussed in section IV.  Finally, the summary is presented in Section V.

\section{Statistics dependence of probability distributions of measured moments}

It is well known that the probability distribution of the measurement should be a normal one in the case of the sufficient statistics. For the statistics at current heavy ion collisions, usual observables, such as,  $K/\pi$ fluctuation~\cite{STAR-kpi}, and the mean of various multiplicity, whose probability distributions are already  normally distributed.

However, for high-order moments, in particular, the fourth, and even the sixth order moments, the required statistics should be largely increased. How much they should be is not yet clear. So in this section we firstly demonstrate how they change with statistics from a simple simulation.

Let's suppose that the particle and anti-particle are produced independently by a Poisson distribution. Then the net-particle follows a Skellam distribution. The parameters of Skellam distributions are the means of particle and anti-particle~\cite{Skellam-1, Skellam-2, Skellam-3}. In our simulation, they are set to $m_1= 14.5$, $m_2 = 0.6$, reference to the
means of proton and anti-proton at low energies of RHIC/STAR~\cite{STAR-proton}.

\begin{figure}[htmp]
\centering
\includegraphics[width=5.5in]{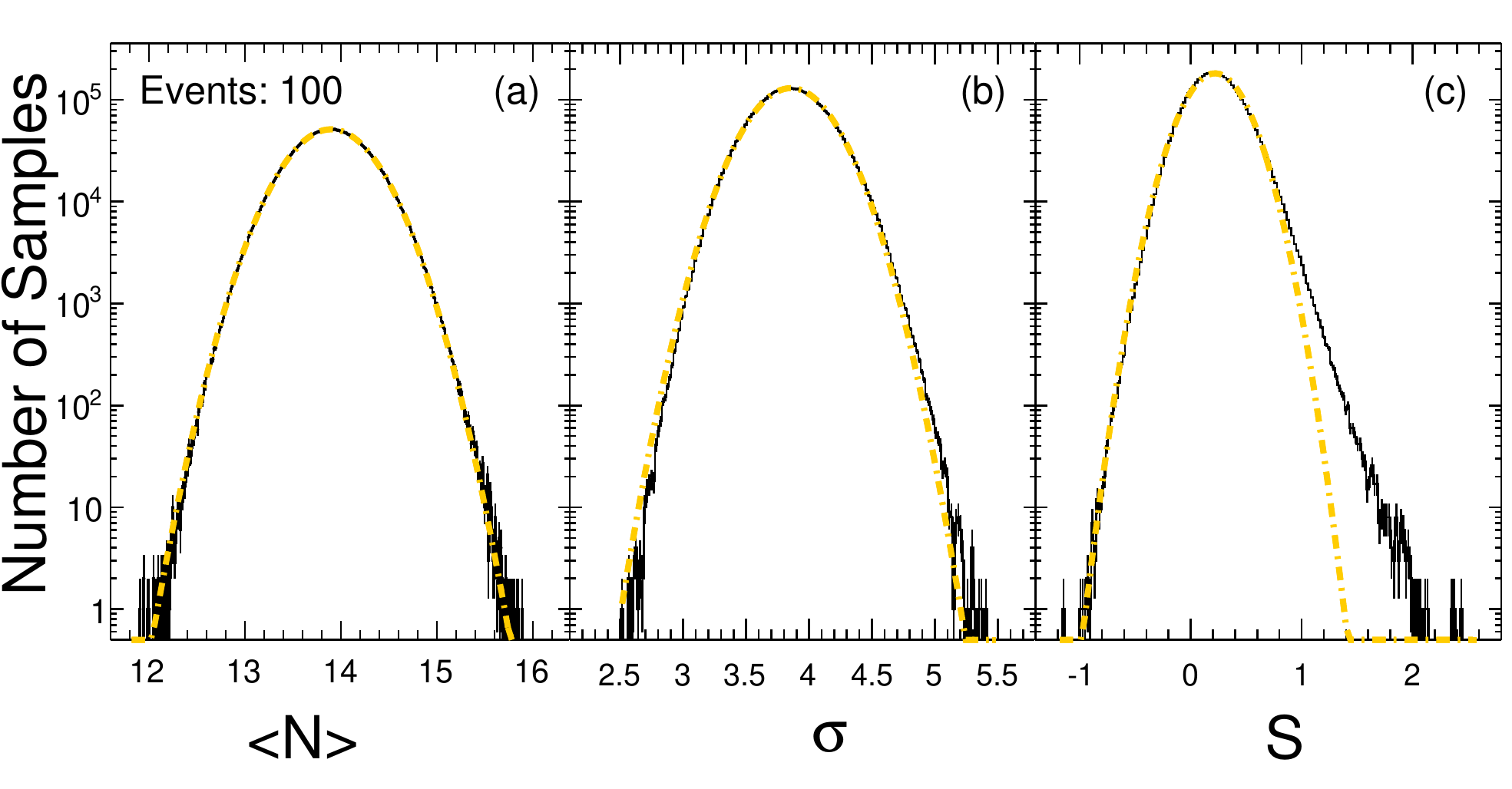}
\caption{\label{moment-mean} (Color online) Probability distributions of measured mean $\left<N\right>$, variance $\sigma$, and skewness $S$ of 100 events produced by Skellam distribution (black points). The yellow dashed lines are corresponding normal fittings.\label{event-100}}
\end{figure}

In Figure~\ref{event-100}, the  distribution of mean $\left<N\right>$, variance $\sigma$, and skewness $S$ of 100 events of Skellan distribution are shown. The distributions are obtained from 5 million (5M) independent samples and presented by black points in Figure~\ref{event-100}(a), (b) and (c), respectively. The yellow dashed lines are fitted by a normal distribution.

We can see from Figure~\ref{event-100}(a) that the distribution of measured mean $\left<N\right>$ is
already a normal one. Its fit goodness $\chi^2/ndf$ is 1.48 as listed in Table~\ref{chi-square}. So the statistics 100 seems enough for first order moment analysis. While, the distribution of the second order moment $\sigma$ in Figure~\ref{event-100}(b) is slightly deviated from normal distribution, and its $\chi^2/ndf=46.3$. The distribution of third order moment $S$ in Figure~\ref{event-100}(c) can not be fitted by a normal one any more with $\chi^2/ndf=228.9$. So the higher the order of moments, the larger statistics are required.

\begin{table}[htmp]
\centering
\begin {tabular}{|c|c|c|c|c|c|} \hline
Statistics  & 100  &  1000 &  10000 & 100000 & 1M\\
\hline
$\chi^2/ndf$ ($Mean$)  & 1.48   & 1.08 & 1.05 & 0.99 & 1.00\\
\hline
$\chi^2/ndf$ ($\sigma$)  & 46.3   & 4.96 & 1.55 & 0.97 & 0.95\\
\hline
$\chi^2/ndf$ ($S$)  & 228.9   & 56.9 &  7.97  & 2.08 & 0.96\\
\hline
$\chi^2/ndf$ ($\kappa$) & 4962 & 1955  &  321 & 37.6 & 4.00\\
\hline
\end{tabular}
\caption{$\chi^2/ndf$ of normally fit for probability distributions of $\left<N\right>$, $\sigma$, $S$ and $\kappa$ with statistics 100, 1000, 10000, 100000 and 1M, respectively. \label{chi-square}}
\end{table}

\begin{figure}[htmp]
\centering
\includegraphics[width=3.4in]{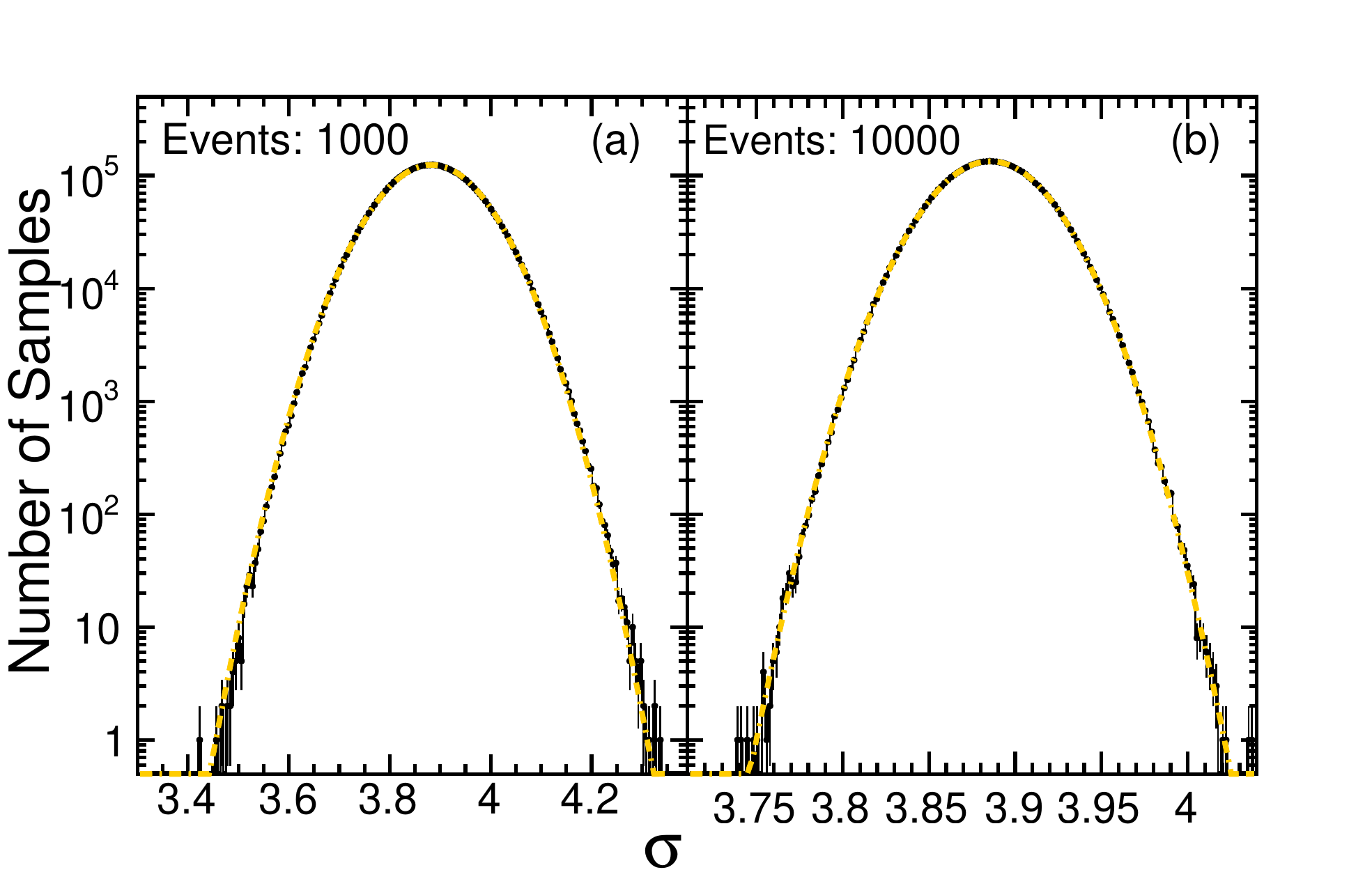}
\includegraphics[width=3.4in]{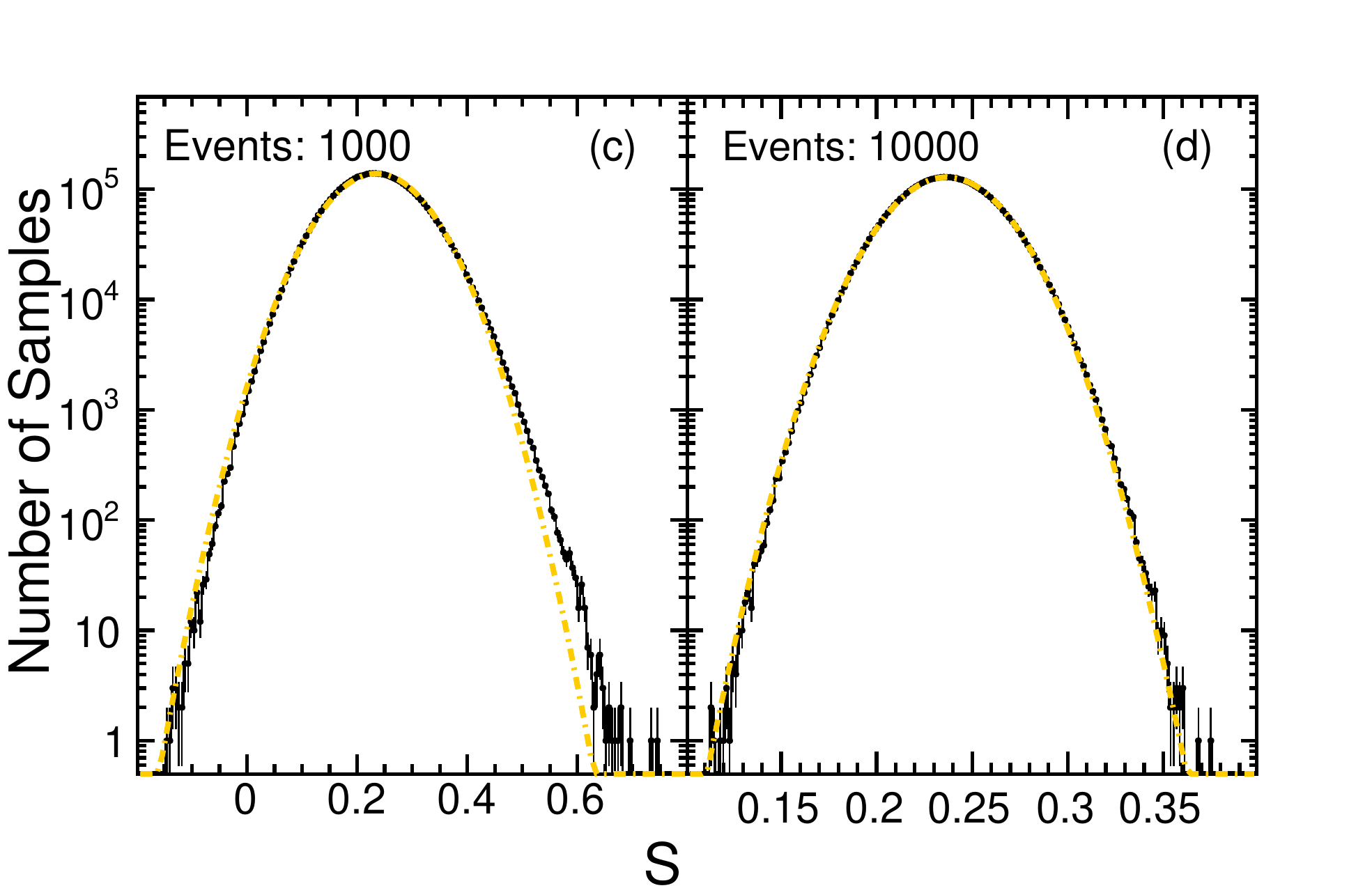}
\caption{\label{moment-2to3} (Color online) Probability distributions of the measured $\sigma$ (upper panel), and $S$ (lower panel) with statistics: 1,000 (left column), and 10,000 (right column), respectively. The yellow dashed lines are the corresponding normal fittings.}
\end{figure}

Increasing the statistics to magnitude 1,000, and 10,000, we re-calculate the second and third
order moments. Their probability distributions are presented in the upper and lower panel of Figure~\ref{moment-2to3}, respectively. The left and right columns correspond to the statistics 1,000, and 10000, respectively.  From Figure~\ref{moment-2to3}(a) with statistics 1,000, we can see that the distribution of $\sigma$ can be fitted by a normal one with $\chi^2/ndf=4.96$.  And Figure~\ref{moment-2to3}(b) with statistics 10,000  gives an  better normal distribution with $\chi^2/ndf=1.55$.

 Therefore, it seems that we can relax the usual requirement of fit goodness from $\chi^2/ndf \sim 1$ to $\chi^2/ndf \sim 10$ in our simulation.  In this case, if the distribution of measured moments is fitted by a normal one with fit goodness $\chi^2/ndf < 10$, the corresponding statistics  is sufficient for a reliable analysis. By this standard, the statistics 1000 seems enough for the second order moment analysis. It is a magnitude larger than that of the first order moment.

Figure~\ref{moment-2to3}(c) shows the probability distribution of the third order moments with statistics 1000. It can still not be well described by a normal one. The fit goodness is $\chi^2/ndf=56.9$, which is still larger than 10. When statistics is increased to 10,000, i.e., Figure~\ref{moment-2to3}(d), it can be fitted by a normal one with $\chi^2/ndf=7.97$. So the statistics for the third moments are 10,000. It is again a magnitude larger than that of the second moment.

Obviously, the statistics for fourth order moment, $\kappa$, has to be larger than 10,000.
From  Table~\ref{chi-square}, we can see that when statistics increases to 1 million (1M),  its fit is goodness $\chi^2/ndf=4.00$. So the statistics for the  fourth order
moment is  around 1M.

\begin{figure}[htmp]
\centering
\includegraphics[width=5.5in]{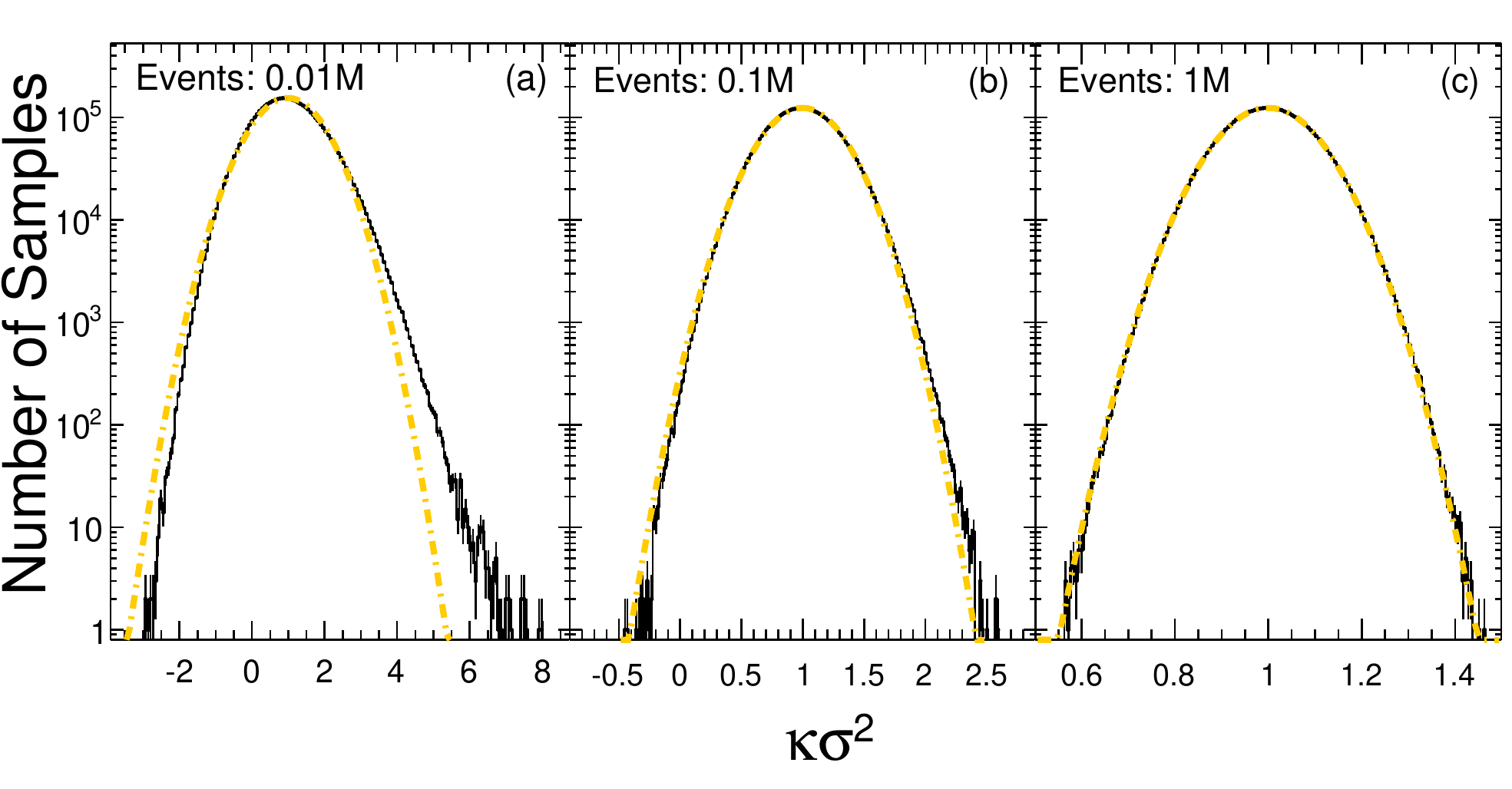}
\caption{\label{moment-ksigma} (Color online)  Probability distributions of $\kappa \sigma^2$.}
\end{figure}

The product of the  fourth and second order moment, $\kappa\sigma^2$, is another interested observable~\cite{Stephanov-PRL-1}. Its statistics should be mainly determined by the fourth order moment $\kappa$. The probability distributions of measured product with statistics 0.01M, 0.1M and 1M are presented in Figure~\ref{moment-ksigma}(a), (b) and (c), respectively. The yellow dashed lines are fitted normal distributions. It shows clearly that figure~3(a) can not be fitted by a normal distribution, and Figure~\ref{moment-ksigma}(b) is still slightly deviated from it.
Only Figure~\ref{moment-ksigma}(c) with statistics 1M can be fitted by a normal one. Consequently, the statistics of $\kappa\sigma^2$ are 1M, the same magnitude as those of $\kappa$.

\section{Statistics dependence of the means of measured moments}

It is known when the statistics is large enough, the mean of measured moments should be saturated
and independent of the statistics, i.e.,
\begin{equation}\label{eq1}
\left<X\right>_{n_1} = \left<X\right>_{n_2} = \left<X\right>_{n_3},
\end{equation}
where $X$ is the measured moments of $\left<N\right>$, $\sigma$, $S$ and $\kappa$. The subscripts $n_1$, $n_2$ and $n_3$ denote different statistics.

In previous study~\cite{statistics-lizhu}, we defined the difference of the mean of measured moments to its expectation is:
\begin{equation}
d=\frac{\left|\left<X\right>_{exp} - \left<X\right>_{th}\right|}{\left|\left<X\right>_{th}\right|}\times100\%,
\end{equation}
where $\left<X\right>_{th}$ is theoretical expectation and $\left<X\right>_{exp}$ is experimentally measured mean of moments with finite statistics. Consider the statistics at RHIC/BES I and its uncertainty of the measurements, the result is acceptable if $d<5\%$~\cite{statistics-lizhu}.
The statistics at RHIC/BES II has a few hundred millions, which is 1-2 magnitudes larger than those at RHIC/BES I. So the statistical uncertainty will be largely reduced. In this case,  $d<1\%$  is more reasonable within smaller uncertainties at planning RHIC/BES II.

The left panel of Table~\ref{table-d} lists statistics dependence of $\left<\sigma\right>$ and corresponding difference $d$, respectively. It shows that the difference $d$ is already less than $1\%$ when statistics is 100. This statistics is less than 1,000, which is obtained in the above section by requiring a normal distribution with the fit goodness $\chi^2/ndf < 10$. So the normal distribution of the measured moment is a more restrictive requirement for the statistics.

\begin{table}[htmp]
\centering
\begin {tabular}{|c|c|c|c|} \hline
Statistics  & 100  &  500 &  1000\\
\hline
$\left<\sigma\right>$  & 3.85629   & 3.87996 & 3.88286 \\
\hline
$d(\%)$  & 0.76   & 0.15 &  0.077 \\
\hline
\end{tabular}
\begin {tabular}{|c|c|c|c|c|} \hline
Statistics  & 1000  &  2000 &  5000 & 10000\\
\hline
$\left<\kappa\sigma^2\right>$  & 0.893   & 0.946 & 0.979 & 0.990 \\
\hline
$d(\%)$  & 10.7   & 5.45&  2.15 & 0.97\\
\hline
\end{tabular}
\caption{Statistics dependence of $\left<\sigma\right>$ and $\left<\kappa\sigma^2\right>$,  and the difference to their saturated value ($d$). \label{table-d}}
\end{table}

In Figure~\ref{average-ksigma}(a) and \ref{average-ksigma}(b), the statistics dependence of $\left<\kappa\sigma^2\right>$ for two sets of Skellam parameters: (a) $ m_1= 14.5$ and $m_2 = 0.6$, and (b) $m_1 = 79.0$ and $m_2=65.0$ (reference to the means of charge and anti-charge within RHIC/STAR acceptance~\cite{STAR-charge}) are presented, respectively. With increase of statistics, $\left<\kappa\sigma^2\right>$ in both Figure~\ref{average-ksigma}(a) and Figure~\ref{average-ksigma}(b) increases firstly and then converges to their expectation unity. While, it converges more quickly in Figure~\ref{average-ksigma}(a) than that in Figure~\ref{average-ksigma}(b). So the required statistics is related to  parent distribution.

\begin{figure}[htmp]
\centering
\includegraphics[width=4.5in]{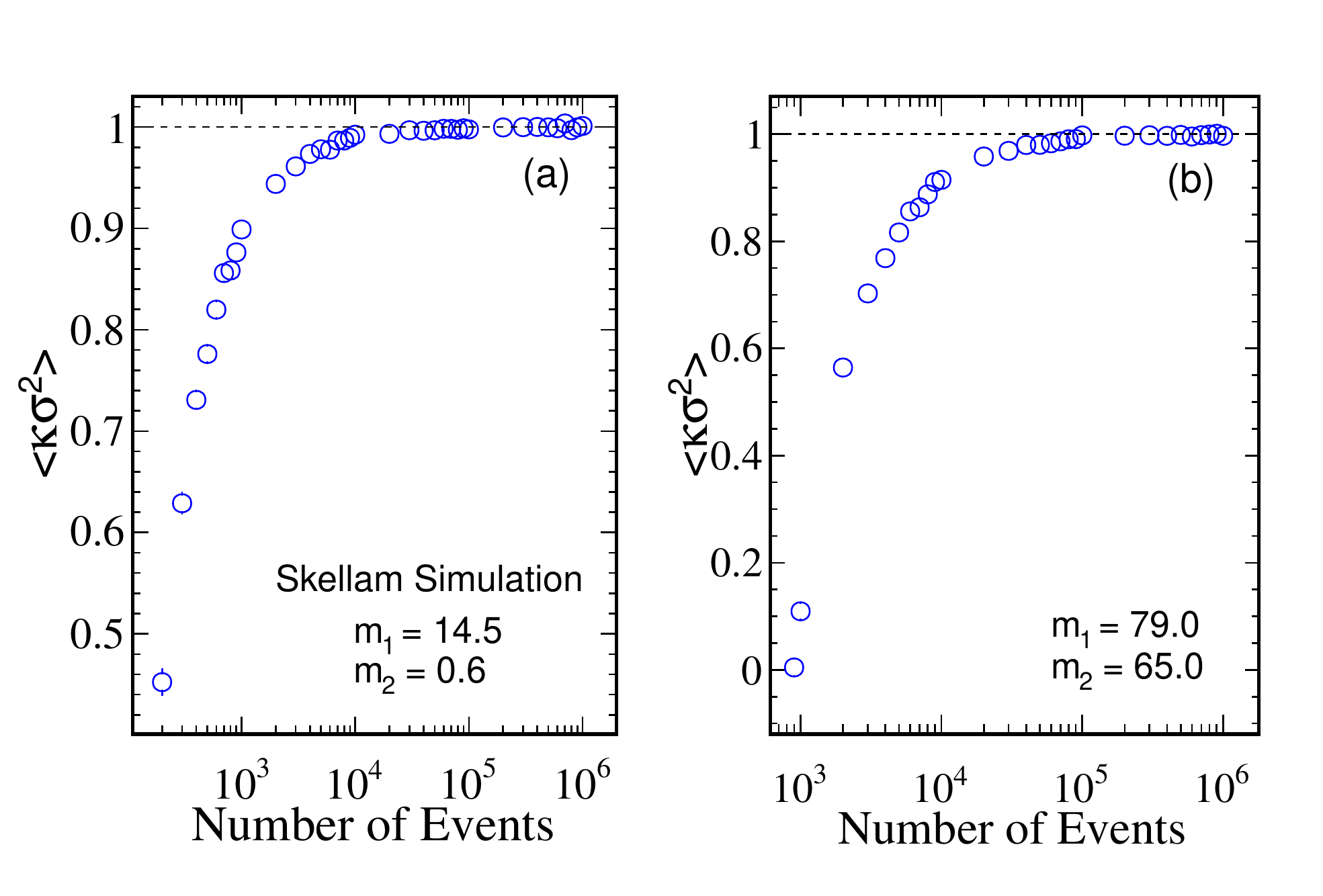}
\caption{\label{average-ksigma}(Color online) Statistics dependence of $\left<\kappa\sigma^2\right>$, which are all averaged from 5M independent samples randomly generated from the simulations of Skellam distributions with the input parameters: (a) $ m_1= 14.5$ and $m_2= 0.6$ and  (b) $m_1= 79$ and $ m_2= 65$, respectively.\label{average-ksigma}}
\end{figure}

The statistics dependence of a few typical $\left<\kappa\sigma^2\right>$ in Figure~\ref{average-ksigma}(a) is given in the right panel of Table~\ref{table-d}. It shows that the difference $d$ is less than 1\%, when statistics is above 10,000. It is about two magnitudes smaller than that obtained from the above section. So the normal distribution of the measured moment is also more restrictive for high-order moments.

\begin{figure}[htmp]
\centering
\includegraphics[width=6.5in]{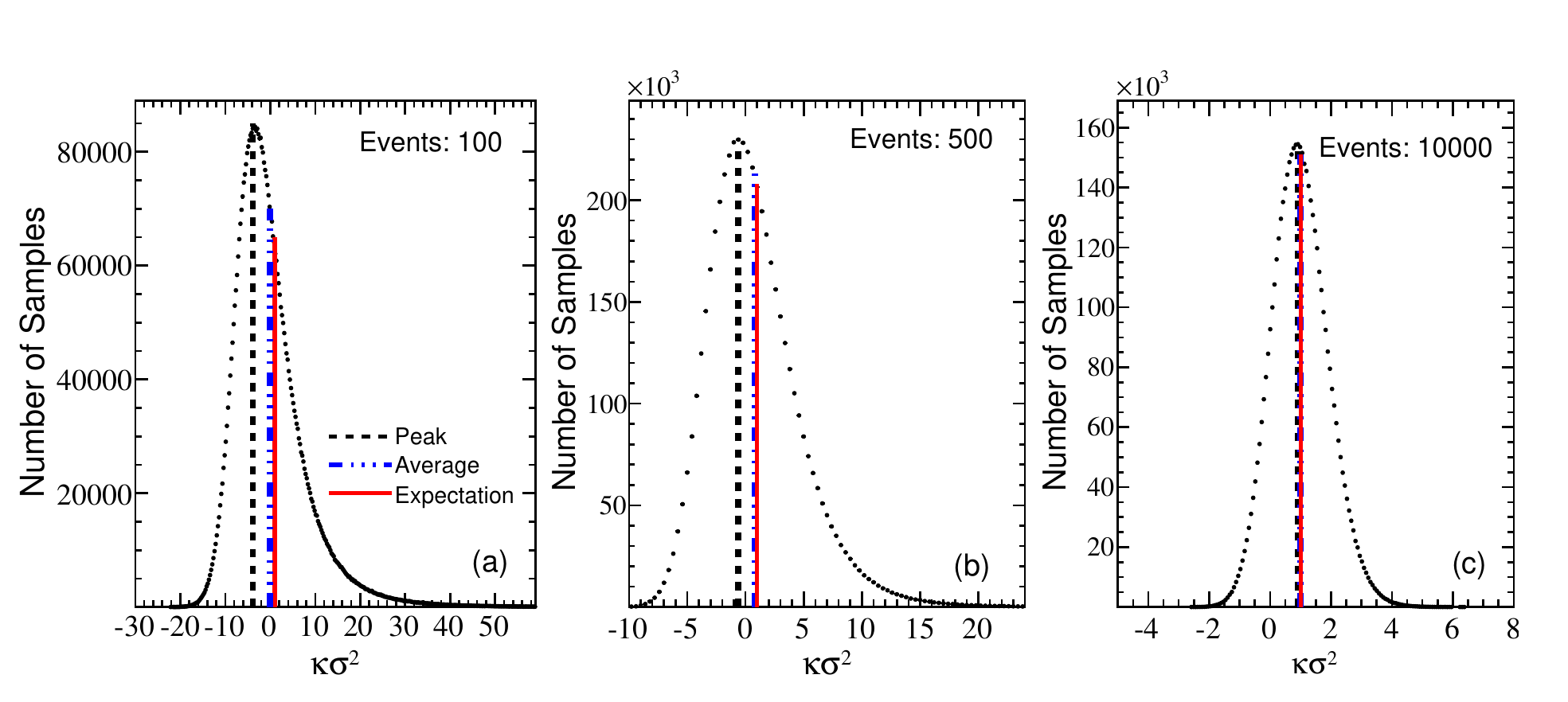}
\caption{\label{ksigma_truncation} (Color online) Probability distributions of $\kappa \sigma^2$ with statistics 100, 500 and  10000 shown from panel (a) to (c), respectively. }
\end{figure}

Figure~\ref{average-ksigma}(a) and \ref{average-ksigma}(b) demonstrate that the mean of measured moments are underestimated,
 when statistics is insufficient. The poorer the statistics, the smaller
the mean. To search for the reasons, the probability distributions of measured $\kappa\sigma^2$ at statistics 100, 500 and 10000 are presented in Figure~\ref{ksigma_truncation}(a), \ref{ksigma_truncation}(b), and \ref{ksigma_truncation}(c), respectively. The black dashed lines shows the peak of the distributions. The blue lines are the mean of their probability distribution, i.e., $\left<\kappa\sigma^2\right>$. The red solid lines are theoretical expectation.

\begin{figure}[htmp]
\centering
\includegraphics[width=6.5in]{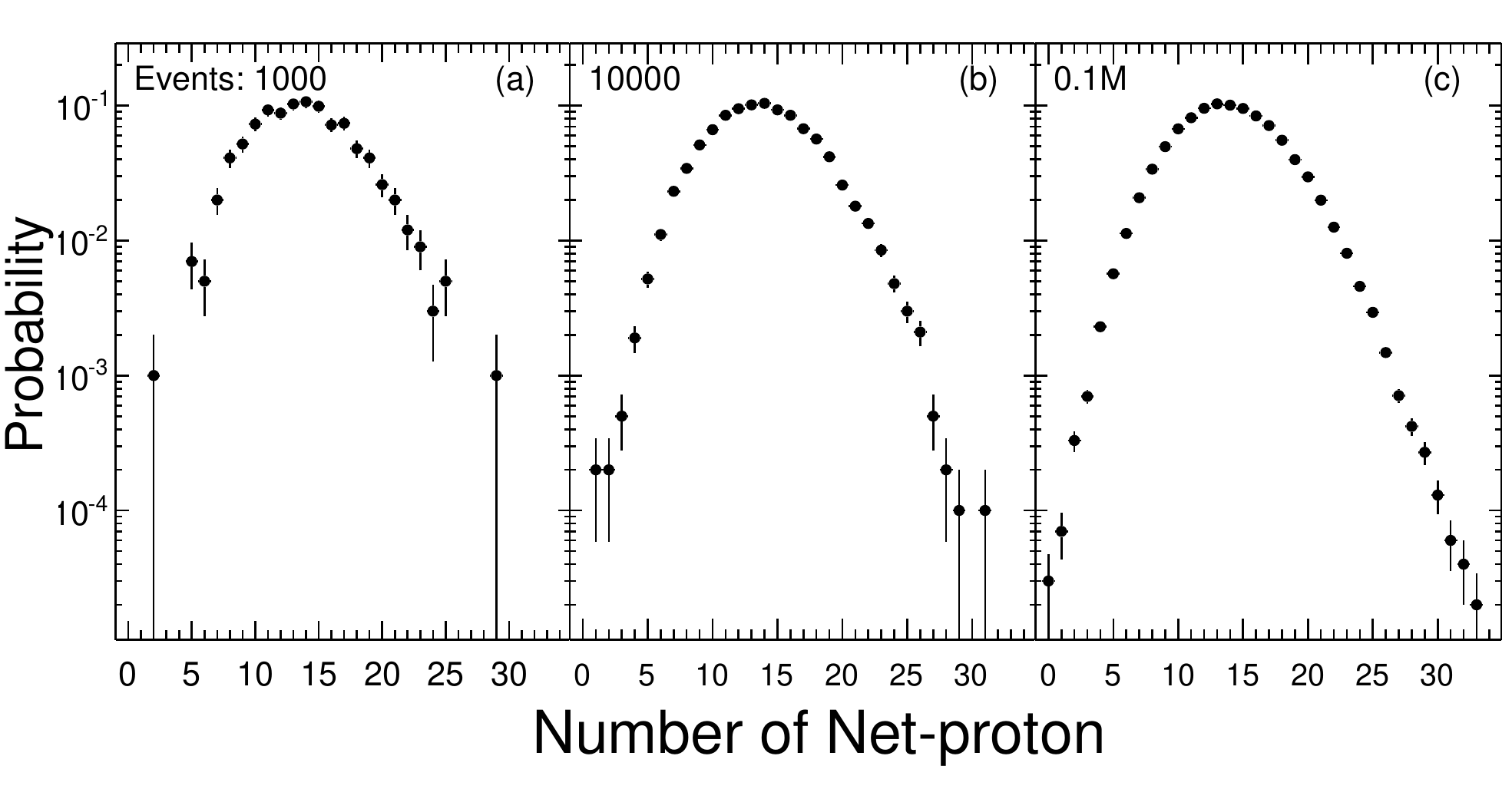}
\caption{\label{net-p-dis}  Net-particle multiplicity distributions with statistics 1,000, 10,000 and 0.1M shown from panel (a) to (c), respectively. }
\end{figure}

Figure~\ref{ksigma_truncation}(a) shows that the peak(black dashed line) and mean(blue line) of the distribution are on the left side of the expectation (unity). The probability on the left side of unity is systematically higher than that on the right side. So the measured moment is underestimated when the statistics is poor.
This is due to the lost information of the tail of net-particle multiplicity distributions, which can be found from Figure~\ref{net-p-dis}.  Figure~\ref{net-p-dis}(a), (b) and (c) show  the net-particle multiplicity distributions with statistics 1,000, 10,000, and 0.1M. These three samples are randomly selected from the sub-samples using for calculating moments in this paper. It shows that the tail of the net-particle multiplicity distributions is  dependent on the statistics. With insufficient statistics, the tail of net-particle multiplicity distributions shown in Figure~\ref{net-p-dis}(a) and (b) can not be well detected. The smaller the statistics, the
poorer information about the tail of the net-particle multiplicity distributions we can
detect.   With the statistics increased, this information can be better shown up, such as shown in Figure~\ref{net-p-dis}(c). That is why $\left<\kappa\sigma^2\right>$ decreases as the statistics decreases in case of insufficient
statistics shown in Figure~\ref{average-ksigma}.

The influence of tail of net-particle multiplicity distributions on measured moments  is similar to what had been observed at measurement of sixth order moment, where the tail of the net-proton multiplicity distributions is artificially truncated~\cite{C6-STAR}. It also shows that the more the lost information at the tail of the distribution, the smaller $C_6/C_2$ is measured.

If the statistics is sufficient, the peak, mean and expectation, i.e., the three lines in Figure~\ref{ksigma_truncation}, should be identical. So we can see from Figure~\ref{ksigma_truncation}(a) to~\ref{ksigma_truncation}(b), and finally to~\ref{ksigma_truncation}(c) that they become closer and closer.

In addition, the statistical uncertainty of Figure~\ref{ksigma_truncation}, i.e., the widths of probability distributions, look much larger than the deviation of the average (blue dashed lines in Figure~\ref{ksigma_truncation}) from the expectation (red solid lines). However, it is the statistical uncertainty of one sub-sample, similar to the experimental measurement at a fixed multiplicity.
An experimental point at given centrality is averaged over all multiplicities in the centrality, analogous to the case of Figure 4. The corresponding statistical uncertainty is usually smaller than those given in Figure 5.
Moreover, it has been shown in Figure 1 in Ref.~\cite{statistics-lizhu} that the fluctuation of many sub-samples are out of $3\sigma$ of expectation. So the deviation of average from its
expectation due to insufficient statistics can not be ignored.

\section{Measurement of $\kappa\sigma^2$  at RHIC/BES I and II}

Up to now, we only consider the required statistics for a given distribution. In fact, this is  corresponding to statistics in each of $N_{ch}$ with CBWC method, or in each  $\delta1\%$ centrality bin with improved CBWC method. The statistics of real sample in minimum bias has to be quite larger than it.
For example, the required statistics for  $\left<\kappa\sigma^2\right>$ of  net-charge is about 0.1M (cf. Figure~\ref{average-ksigma} and Table~\ref{table-d}), while the
corresponding statistics in whole minimum bias sample is around 200M assuming there is 100 multiplicity bin in 0-5\% centrality~\cite{STAR-centrality}. That is why we propose the improved CBWC method on moment analysis.

As we have mentioned in motivation, another advantage of the improved CBWC method with each $\delta1\%$ centrality bin is that we can further investigate why $\kappa\sigma^2$ in central collisions is systematically lower than the other centralities. With  tens of millions of statistics at RHIC/BES I, the number of events in each $\delta1\%$ is around 0.1M, similar statistics as  Figure~\ref{moment-ksigma}(b).
If we study behavior of $\kappa\sigma^2$ in each $\delta1\%$ centrality bin, the statistical uncertainties shown in  Figure~\ref{moment-ksigma}(b)  is around 0.29. Consequently, the statistical uncertainties has already larger than the measured difference of centrality dependence of  $\kappa\sigma^2$ shown in Ref.~\cite{STAR-proton}.


At RHIC/BES II, the statistics in minimum bias will be increased to 100-400 millions, cf. Table~\ref{statistics-BESII}.
 The statistics in suggested $\delta1\%$ centrality bin will be about 1-4 millions. The width in Figure~\ref{moment-ksigma}(c) is about 0.1, which would be slightly larger than statistical uncertainties at RHIC/BES II.
 So in this case, the statistics should be sufficient for $\kappa\sigma^2$ analysis in each $\delta1\%$ centrality bin.  This will be helpful to clarify why measured $\kappa\sigma^2$ of net-proton multiplicity distributions in central collisions (0-5\% and 5-10\%) is much lower than those in the other centralities.


\section{Summary}

In summary, the moments of conserved quantities are important measurements to study QCD phase structure and bulk properties. If the statistics is insufficient, the shape of net-proton/net-charge multiplicity distributions, especially for the tail of the distributions, can not be well detected. Consequently, the measured moments are underestimated and the obtained moments probability distributions would be deviated to normal distribution.

In this paper, statistics dependence of the probability distributions and the means of measured moments and their product $\kappa\sigma^2$ are systematically presented.
 The behavior of simulation in this paper can be referenced to moments of net-proton multiplicity distributions at RHIC low energies.
It is found that when the probability distribution of the measured moment can be fitted by a normal one with fit goodness $\chi^2/ndf < 10$, the corresponding statistics is enough for the moment analysis.
The required statistics can also be obtained when the mean of the measured moment is saturated to its expectation within statistical error. By this way, the required statistics is 1-2 magnitude less than those obtained from probability distribution of measured moments. So the normal
distribution of the measured moment is a more restrictive requirement for the statistics.

 Currently, the CBWC method in each of $N_{ch}$ is  employed to moments analysis. With the restrictive requirement of the normally probability distribution,  it is found that the statistics at RHIC/BES I is enough to extract the second and third order moments. However,  at RHIC low energy, the statistics is not enough  for $\kappa\sigma^2$ either from the restrictive  normally distribution or quite looser requirement obtained from the saturation of the mean of measured moments. The measured moment will be underestimated if statistics are not sufficient. With the improved CBWC method in each $\delta1\%$ centrality bin, we can obtain much reliable $\kappa\sigma^2$ at RHIC low energies.



The current measured $\kappa\sigma^2$ of net-proton multiplicity distributions shows that $\kappa\sigma^2$ in central collisions is systematically lower than that in the other centralities. With planning statistics for RHIC/BES II, and improved CBWC method, it is able to measure the high moments of conserved quantities in finer centrality bin scale. This will
 help us to  understand  the observation of the centrality and energy dependence of high-order moments.

\section{Acknowledgements}

This work is supported in part by the Major State Basic Research Development Program of China under Grant No. 2014CB845402, the NSFC of China under Grants  No. 11405088, No. 11221504, No. 11005046,  No. 11447190, and the Ministry of Education of China with Project No. 20120144110001.

\clearpage
\end{document}